\documentclass[14pt, fleqn]{extarticle}

\usepackage{latexsym}
\usepackage{amsmath}
\usepackage[colorlinks = true,
            linkcolor = blue,
            urlcolor  = blue,
            citecolor = red,
            anchorcolor = blue]{hyperref}
\usepackage{amsfonts}
\usepackage{amssymb}
\usepackage{graphicx}
\usepackage{color}
\usepackage{tikz}
\usepgflibrary{arrows}
\usepackage{mathbbol}
\usepackage{nicefrac, xfrac}
\usepackage[a4paper,bindingoffset=0.2in,
           left=1in,right=1in,top=1in,bottom=1in]{geometry}
\usepackage{enumerate} 
\title{On the generalization\\ of classical Zernike system}
\author{
Cezary Gonera\footnote{cezary.gonera@uni.lodz.pl}, Joanna Gonera\footnote{joanna.gonera@uni.lodz.pl},\\ Piotr Kosi\'nski\footnote {piotr.kosinski@uni.lodz.pl}\\
\small \textit {Faculty of Physics and Applied Informatics}\\
\small \textit {University of Lodz, Lodz, Poland}\\
}

\date{}
\begin{document}
\maketitle
\begin{abstract}
\par We generalize the results obtained recently (Nonlinearity \underline{36} (2023), 1143) by providing a very simple proof of the superintegrability of the Hamiltonian $H=\vec{p}\,^{2}+F(\vec{q}\cdot\vec{p})$, $\vec{q}, \vec{p}\in\mathbb{R}^{2}$, for any analytic function $F$. The additional integral of motion is constructed explicitly and shown to reduce to a polynomial in canonical variables for polynomial $F$. The generalization to the case $\vec{q}, \vec{p}\in \mathbb{R}^{n}$ is sketched.

\end{abstract}

\section{Introduction} 
\label{I}
\par In the recent paper \cite{b1} the natural generalization of the classical Zernike system \cite{b2,b3,b4} (for the quantum version see \cite{b3,b4,b5,b6,b7,b8,b9}) has been proposed. It is defined by the Hamiltonian
\begin {align} 
\label {al1}
H=\vec{p}\,^{2}+\sum^{N}_{n=1}\gamma_{n}(\vec{q}\cdot\vec{p})^{n}
\end {align}
with $\vec{q}=(q_{1},q_{2})$, $\vec{p}=(p_{1},p_{2})$ being canonical variables.
\par It has been shown in \cite{b1} that (\ref{al1}) is maximally superintegrable and admits, apart from energy and angular momentum, an additional integral of motion which is polynomial of degree $N$ in momenta. 
\par In the present paper we offer a very simple proof of superintegrability of a more general Hamiltonian 
\begin {align} 
\label {al2}
H=\vec{p}\,^{2}+F(\vec{q}\cdot\vec{p})
\end {align}
with $F(\cdot)$ being an arbitrary analytic real function. It still possesses two commuting integrals, 
\setlength{\jot}{15pt}
\begin {align} 
\label {al3}
E&=H\nonumber \\
L&=q_{1}p_{2}-q_{2}p_{1}
\end {align}
so again it is sufficient to find the third, functionally independent, integral of motion. We give the explicit form of the latter. It appears to reduce to the polynomial in momenta provided $F(\cdot)$ is a polynomial, in agreement with the results presented in  Ref. \cite{b1}.  Moreover, we show that (\ref{al2}) continues to be superintegrable if the Euclidean plane is replaced by sphere or hyperbolic space. Finally, we extend our results to the case of configuration space of arbitrary dimension.

\section{Superintegrability} 
\label{II}
\par In order to prove the superintegrability of the Hamiltonian (\ref{al2}) let us note that $\vec{q}\cdot \vec{p}$ generates dilatations. Therefore, it Poisson-commutes with any homogeneous function of degree zero. Any such function is a constant of motion for the Hamiltonian (\ref{al2}) provided it is a constant of motion for the free Hamiltonian $H_{0}=\vec{p}\,^{2}$. For example, one can take $Q_{k}(\underline{p})/R_{k}(\underline{p})$ with $Q_{k}$ and $R_{k}$ being homogeneous polynomials of degree $k$. The simplest choice seems to be $\frac{p^{2}_{1}}{\vec{p}\,^{2}}$ or $\frac{p^{2}_{2}}{\vec{p}\,^{2}}$. However, such integrals are singular functions of the momenta while we would like to find the analytic ones.
\par To this end note that $F(\vec{q}\cdot\vec{p})$, being analytic, can be rewritten as
\setlength{\jot}{15pt}
\begin {align} 
\label {al4}
F(\vec{q}\cdot \vec{p})=A\big( (\vec{q}\cdot \vec{p})^{2}\big)+(\vec{q}\cdot \vec{p})B\big( (\vec{q}\cdot \vec{p})^{2}\big)
\end {align}
with
\setlength{\jot}{15pt}
\begin {align} 
\label {al5}
A\big( (\vec{q}\cdot \vec{p})^{2}\big)\equiv \frac{1}{2}\big(F(\vec{q}\cdot \vec{p})+F(-\vec{q}\cdot \vec{p})\big )
\end {align}

\setlength{\jot}{15pt}
\begin {align} 
\label {al6}
B\big( (\vec{q}\cdot \vec{p})^{2}\big)\equiv\frac{1}{2\,\vec{q}\cdot\vec{p}}\Big(F(\vec{q}\cdot \vec{p})-F(-\vec{q}\cdot \vec{p})\Big)
\end {align}
being analytic as well. Due to $(\vec{q}\cdot \vec{p})^{2}=\vec{q}\,^{2}\cdot\vec{p}\,^{2}-L^{2}$ one has

\setlength{\jot}{15pt}
\begin {align} 
\label {al7}
H=\vec{p}\,^{2}+A\,(\vec{q}\,^{2}\cdot\vec{p}\,^{2}-L^{2})+(\vec{q}\cdot \vec{p})\,B\,(\vec{q}\,^{2}\cdot\vec{p}\,^{2}-L^{2})
\end {align}
Finally, note a simple identity
\setlength{\jot}{15pt}
\begin {align} 
\label {al8}
&\frac{p^{2}_{2}}{\vec{p}\,^{2}}\cdot H - \frac{p^{2}_{2}}{\vec{p}\,^{2}}A(-L^{2})-\frac{p_{1}\cdot p_{2}\cdot L}{\vec{p}\,^{2}}B(-L^{2})=\nonumber\\
&= p^{2}_{2}+\frac{p^{2}_{2}}{\vec{p}\,^{2}}\big(A(\vec{q}\,^{2}\cdot \vec{p}\,^{2}-L^{2})-A(-L^{2})\big)+\nonumber\\
&+\frac{p^{2}_{2}}{\vec{p}\,^{2}}\cdot (\vec{q}\cdot \vec{p})\big(B(\vec{q}\,^{2}\cdot \vec{p}\,^{2}-L^{2})-B(-L^{2})\big)+q_{2}p_{2}B(-L^{2})
\end {align}
The left hand side of eq. (\ref{al8}) obviously provides an integral of motion functionally independent of $H$ and $L$ while the right hand side is explicitly analytic. If $F$ is a polynomial it is a polynomial of the same degree. Similar identity holds under the replacement $1\leftrightarrow 2$.

\section{Polar coordinates} 
\label{III}
\par Due to the rotational invariance it is interesting to study the Hamiltonian (\ref{al2}) in polar coordinates. To this end we make the canonical transformation to polar variables \cite{b1}:

\setlength{\jot}{15pt}
\begin {align} 
\label {al9}
q_{1}&=r \cos \varphi & p_{1}&=p_{r}\cos\varphi -\frac{p_{\varphi}}{r}\sin \varphi\nonumber \\
q_{2}&=r \sin \varphi & p_{2}&=p_{r}\sin\varphi +\frac{p_{\varphi}}{r}\cos \varphi
\end {align}
where $p_{\varphi}=L$ is the integral of motion. The Hamiltonian acquires the form
\begin {align} 
\label {al10}
H=p^{2}_{r}+\frac{p^{2}_{\varphi}}{r^{2}}+F(r\cdot p_{r})
\end {align}
while the canonical equations of motion read 
\setlength{\jot}{15pt}
\begin {align} 
\label {al11}
\dot{r}&=\frac{\partial H}{\partial p_{r}}=2p_{r}+F'(r\cdot p_{r})r\nonumber \\
\dot{p}_{r}&=-\frac{\partial H}{\partial r}=\frac{2p^{2}_{\varphi}}{r^{3}}-F'(r\cdot p_{r})p_{r}\nonumber\\
\dot{\varphi}&=\frac{\partial H}{\partial p_{\varphi}}=\frac{2p_{\varphi}}{r^{2}}\nonumber\\
\dot{p}_{\varphi}&=-\frac{\partial H}{\partial \varphi}= 0
\end {align}
\par It is now easy to see that our system is superintegrable. Namely, there exists third, functionally independent, integral of motion: 
\begin {align} 
\label {al12}
C= \text{any periodic function of}\,\bigg( \text{arctan} \Big(\frac{r \cdot p_{r}}{p_{\varphi}}\Big)-\varphi\bigg)
\end {align}
For example, for any natural $n$ one can take
\begin {align} 
\label {al13}
C_{n}\equiv \sin \Bigg( 2n \bigg( \text{arctan} \Big(\frac{r \cdot p_{r}}{p_{\varphi}}\Big)-\varphi\bigg)\Bigg)
\end {align}
An elementary computation yields
\begin {align} 
\label {al14}
C_{n}=\bigg(\frac{z^{2n}-\bar{z}^{2n}}{2i\vert z\vert^{2n}}\bigg)\cos 2 n\varphi -\bigg(\frac{z^{2n}+\bar{z}^{2n}}{2\vert z\vert ^{2n}}\bigg)\sin 2 n \varphi
\end {align}
with
\begin {align} 
\label {al15}
z\equiv p_{\varphi}+ i r\cdot p_{r}
\end {align}
Obviously, there is only one functionally independent integral $C_{n}$. It is convenient to choose the simplest one, 
\begin {align} 
\label {al16}
C_{1}=\bigg (\frac{z^{2}-\bar{z}^{2}}{2i\vert z\vert^{2}}\bigg)\cos 2 \varphi -\bigg(\frac{z^{2}+\bar{z}^{2}}{2\vert z\vert ^{2}}\bigg)\sin 2 \varphi
\end {align}
$H(E)$ and $p_{\phi}(L)$ can be expressed in terms of $z$, $\bar{z}$ and $r$ as follows
\begin {align} 
\label {al17}
E=H=\frac{\vert z \vert ^{2}}{r^{2}}+F\bigg(\frac{z-\bar{z}}{2i}\bigg)
\end {align}
\begin {align} 
\label {al18}
L=p_{\varphi}=\frac{z+\bar{z}}{2}
\end {align}
\par Assume now that $F(\cdot)$ is a polynomial; actually, we shall consider only the case of $F(\cdot)$ being a monomial and leave the general case to the reader. So we put 
\begin {align} 
\label {al19}
F(r\cdot p_{r})=\gamma(r\cdot p_{r})^{N}\quad \text{,}\quad \gamma \in \mathbb{R}
\end {align}
\par We show that the third integral of motion can be chosen as a polynomial in the momenta. The starting point is our integral $C_{1}$, eq. (\ref{al16}). We consider two cases:\\

(i)\;\; N=2k\\
\\
Let us define
\begin {align} 
\label {al20}
G\equiv H-(-1)^{k}\gamma p_{\varphi}^{\phantom{\varphi}2k}
\end {align}
Obviously, $G$ is an integral of motion; in terms of $z$, $\bar{z}$ variables it reads
\begin {align} 
\label {al21}
G=\frac{\vert z\vert^{2}}{r}+\gamma\frac{(-1)^{k}}{2^{2k}}\Big((z-\bar{z})^{2k}-(z+\bar{z})^{2k}\Big)
\end {align}
The second term on the right hand side is proportional to $\vert z \vert ^{2}$ (the terms $z^{2k}$ and $\bar{z}^{2k}$ cancel). Therefore, the integral of motion
\begin {align} 
\label {al22}
\widetilde{C}\equiv G \cdot C_{1}
\end {align}
is a polynomial in $z$, $\bar{z}$ of degree $N$ which implies that $\widetilde{C}$ is a polynomial in $p_{r}$ and $p_{\varphi}$ (consequently, in $p_{1}$ and $p_{2}$) of the same degree.
\\

(ii)\;\;N=2k+1\\
\par In this case the previous argument does not work unless $\gamma$ is purely imaginary. However, one can proceed as follows. Note that, given any $\chi=\chi (E,L)$, $G=G(E,L)$, the following function is an integral of motion
\begin {align} 
\label {al23}
\widetilde{C}_{\chi}\equiv G\bigg(\frac{z^{2}-\bar{z}^{2}}{2i\vert z \vert ^{2}} \bigg)\cos (2\varphi +\chi)-G\bigg(\frac{z^{2}+\bar{z}^{2}}{2\vert z \vert ^{2}} \bigg)\sin (2\varphi+\chi)
\end {align}
which can be rewritten as
\begin {align} 
\label {al24}
\widetilde{C}_{\chi}=A\cos 2 \varphi-B\sin 2 \varphi
\end {align}
\begin {align} 
\label {al25}
A\equiv G\bigg(\frac{z^{2}-\bar{z}^{2}}{2i\vert z \vert ^{2}} \bigg)\cos\chi-G\bigg(\frac{z^{2}+\bar{z}^{2}}{2\vert z \vert ^{2}} \bigg)\sin\chi
\end {align}
\begin {align} 
\label {al26}
B\equiv G\bigg(\frac{z^{2}-\bar{z}^{2}}{2i\vert z \vert ^{2}} \bigg)\sin\chi+G\bigg(\frac{z^{2}+\bar{z}^{2}}{2\vert z \vert ^{2}} \bigg)\cos\chi
\end {align}
Now, in order to $\widetilde{C}_{\chi}$ become a polynomial in $z$ and $\bar{z}$, $G$ and $\chi$ should be chosen in such a way that both 
\setlength{\jot}{15pt}
\begin {align} 
\label {al27}
\frac{(z^{2}-\bar{z}^{2})}{i}G\cos\chi-(z^{2}+\bar{z}^{2})G\sin\chi\nonumber \\
\frac{(z^{2}-\bar{z}^{2})}{i}G\sin\chi+(z^{2}+\bar{z}^{2})G\cos\chi
\end {align}
are proportional to $\vert z \vert^{2}$. Taking the appropriate linear combinations of the above expressions we finally conclude that 
\begin {align} 
\label {al28}
G\cos\chi-i\,G\sin\chi
\end {align}
should be proportional to $\bar{z}$. Now, 
\begin {align} 
\label {al29}
E=\frac{\vert z \vert ^{2}}{r^{2}}+i\,\gamma\,(-1)^{k+1}\,\frac{(z-\bar{z})^{2k+1}}{2^{2k+1}}
\end {align}
Therefore, one can take
\begin {align} 
\label {al30}
G \sin \chi=E\quad \text{,}\quad G\cos \chi =(-1)^{k}\,\gamma\, p_{\varphi}^{2k+1}
\end {align}
With this choice $\widetilde{C}_{\chi}$ becomes a polynomial in $z$ and $\bar{z}$ of degree $N=2k+1$.

\section{Simple examples} 
\label{IV}
Consider the classical Zernike system,
\begin {align} 
\label {al31}
H=p^{2}_{r}+\frac{p^{2}_{\varphi}}{r^{2}}+\gamma_{1}(r\cdot p_{r})+\gamma_{2}(r\cdot p_{r})^{2}
\end {align}
or
\begin {align} 
\label {al32}
H=\frac{\vert z \vert^{2}}{r^{2}}+\frac{\gamma_{1}}{2i}(z-\bar{z})-\frac{\gamma_{2}}{4}(z-\bar{z})^{2}\quad\text{;}
\end {align}
here $F(\cdot)$ is no longer a monomial. However, one easily checks that the choice 
\setlength{\jot}{15pt}
\begin {align} 
\label {al33}
G\cos\chi&=\gamma_{1}\,p_{\varphi}\nonumber \\
G\sin\chi&=E+\gamma_{2}\,p^{2}_{\varphi}
\end {align}
yields, through eqs. (\ref{al19})-(\ref{al21}), the integral $I'-I$ from Ref. \cite{b1} (cf. eqs. (2.2) therein).
\par Let us now consider the monomial of third order, $N=3\; (k=1)$. Then
\begin {align} 
\label {al34}
E=\frac{\vert z \vert^{2}}{r^{2}}+i\,\gamma\,\frac{(z-\bar{z})^{3}}{2^{3}}
\end {align}
Using eq. (\ref{al30}) one can check that our integral $\widetilde{C}_{\chi}$ takes the form
\setlength{\jot}{15pt}
\begin {align} 
\label {al35}
\widetilde{C}_{\chi}&=\Bigg(\bigg(p^{2}_{r}-\frac{p^{2}_{\varphi}}{r^{2}}\bigg)+\gamma\,\bigg(r^{3}p^{3}_{r}-2r\cdot p_{r}p^{2}_{\varphi}\bigg)\Bigg)\cos 2 \varphi+\nonumber \\
&-\bigg(\frac{2p_{r}p_{\varphi}}{r}+\gamma\,\bigg(2r^{2}p^{2}_{r}p_{\varphi}-p^{3}_{\varphi}\bigg)\Bigg)\sin 2 \varphi
\end {align}
which agrees with the result following from Table 2 of Ref. \cite{b1}.
\par Let us note that obviously one can choose the additional integral of motion as an element of arbitrary fixed representation of the $SO(2)$ rotation group generated by $p_{\varphi}$. Eq. (\ref{al30}) tells us that this can be always real irreducible sum of two complex representations corresponding to the characters $\pm 2$.

\section{Conclusions} 
\label{V}
\par We have shown that the Hamiltonian (\ref{al2}) is maximally superintegrable for any choice of the function $F(\cdot)$. For $F$ being a polynomial it is easy to show that the additional integral of motion can be also chosen to be a polynomial in momenta of the same degree. Moreover, it is an element of the representation of $SO(2)$ group being the direct sum (in complex domain) of representations described by the characters $\pm 2$. In other words, when the complicated polynomial in Cartesian coordinates, representing the relevant integral, is expressed in terms of polar ones, all trigonometric functions cancel except $\sin 2\varphi$ and $\cos 2 \varphi$.
\par We considered only the Euclidean case. However, it is immediate to see that one can proceed analogously if the Euclidean plane is replaced by sphere or hyperbolic space. The only difference is that now the additional integral of motion is any periodic function of 
\begin {align} 
\label {al36}
\text{arctan}\bigg(\frac{T_{k}(\rho)\cdot p_{\rho}}{p_{\varphi}}\bigg)-\varphi
\end {align}
with the notation adopted from Ref. \cite{b1}. Finally, let us note that the Hamiltonian (\ref{al2}) continues to be maximally superintegrable in any dimension, $(\vec{q},\vec{p})\in \mathbb{R}^{2n}$. As an example consider $n=3$. Passing to the spherical coordinates we find
\begin {align} 
\label {al37}
H=p^{2}_{r}+\frac{\vec{L}\,^{2}}{r^{2}}+F(r\cdot p_{r})
\end {align}
\par Now, the motion on $S^{2}$ as configuration space, defined by the Hamiltonian $\vec{L}^{2}$, is maximally superintegrable. Let $(I_{1},I_{2},\psi_{1},\psi_{2})$ be the corresponding action-angle variables; one can take $I_{1}=L_{3}\equiv p_{\varphi}$, $I_{2}=\vert\vec{L}\vert$. Then $I_{1}, I_{2}$ and any periodic function of $\psi_{1}$ are independent integrals of motion. The fourth one is provided by the Hamiltonian (\ref{al37}). The latter can be rewritten as 
\begin {align} 
\label {al38}
H=p^{2}_{r}+\frac{I^{2}_{2}}{r^{2}}+F(r\cdot p_{r})
\end {align}
Repeating the reasoning presented in Sec. III we find the fifth integral 
\begin {align} 
\label {al39}
C=\text{any periodic function of}\; \bigg(\text{arctan}\Big(\frac{r\cdot p_{r}}{I_{2}}\Big)-\psi_{2}\bigg)
\end {align}
\par For arbitrary $n$ we note that the geodesic motion on $S^{n-1}$ is again maximally superintegrable yielding $2(n-1)-1$ integrals of motion. Together with total energy we have $2n-2$ integrals. The remaining one can be constructed along the lines sketched above. 
\\
\\
{\bf Acknowledgments}
\\
We are grateful to Prof. Prof. Krzysztof Andrzejewski and Pawe\l{} Ma\'slanka for helpful discussions and useful suggestions. This paper was supported by the IDUB grant, Decision No 54/2021.

\end {document}